\documentclass[10pt,conference]{IEEEtran}

\usepackage{color}
\usepackage{subfigure}

\usepackage[latin1]{inputenc}
\usepackage[pdftex]{graphicx}

\usepackage{amsmath}
\usepackage{amssymb}
\usepackage{amsfonts}

\IEEEoverridecommandlockouts

\begin{document}

\title{A Cooperative Network Coding Strategy for the Interference Relay Channel}

% ----- Full -----
%\author{\IEEEauthorblockN{Huyen-Chi Bui\IEEEauthorrefmark{1}\IEEEauthorrefmark{2}\IEEEauthorrefmark{3},
%Hugo M{\'e}ric\IEEEauthorrefmark{1}\IEEEauthorrefmark{3},
%J{\'e}r{\^o}me Lacan\IEEEauthorrefmark{1}\IEEEauthorrefmark{3} and
%Marie-Laure Boucheret\IEEEauthorrefmark{2}}

%\IEEEauthorblockA{\IEEEauthorrefmark{1}Universit{\'e} de Toulouse, ISAE/DMIA, France}
%\IEEEauthorblockA{\IEEEauthorrefmark{2}Universit{\'e} de Toulouse, IRIT/ENSEEIHT, France}
%\IEEEauthorblockA{\IEEEauthorrefmark{3}T{\'e}SA, Toulouse, France\\
%Email: \{huyen-chi.bui, hugo.meric, jerome.lacan\}@isae.fr, marie-laure.boucheret@enseeiht.fr}}

\author{\IEEEauthorblockN{Huyen-Chi Bui\IEEEauthorrefmark{1}\IEEEauthorrefmark{2},
Hugo M{\'e}ric\IEEEauthorrefmark{1}\IEEEauthorrefmark{2},
J{\'e}r{\^o}me Lacan\IEEEauthorrefmark{1}\IEEEauthorrefmark{2} and
Marie-Laure Boucheret\IEEEauthorrefmark{1}\IEEEauthorrefmark{2}}

\IEEEauthorblockA{\IEEEauthorrefmark{1}Universit{\'e} de Toulouse, France}
\IEEEauthorblockA{\IEEEauthorrefmark{2}T{\'e}SA, Toulouse, France\\
Email: \{huyen-chi.bui, hugo.meric, jerome.lacan\}@isae.fr, marie-laure.boucheret@enseeiht.fr}}

% ----- All on one line -----
%\IEEEauthorblockA{\IEEEauthorrefmark{1}Universit{\'e} de Toulouse, ISAE/DMIA, France, \IEEEauthorrefmark{2}Universit{\'e} de Toulouse, IRIT/ENSEEIHT, France}
%, \IEEEauthorrefmark{3}T{\'e}SA, Toulouse, France\\
%Email: \{huyen-chi.bui, hugo.meric, jerome.lacan\}@isae.fr, marie-laure.boucheret@enseeiht.fr}}
%\IEEEauthorblockA{\IEEEauthorrefmark{3}T{\'e}SA, Toulouse, France\\
%Email: \{huyen-chi.bui, hugo.meric, jerome.lacan\}@isae.fr, marie-laure.boucheret@enseeiht.fr}}

% make the title area
\maketitle

\begin{abstract}
In this paper, we study an interference relay network with a satellite as relay. We propose a cooperative strategy based on physical layer network coding and superposition modulation decoding for uni-directional communications among users. The performance of our solution in terms of throughput is evaluated through capacity analysis and simulations that include practical constraints such as the lack of synchronization in time and frequency. We obtain a significant throughput gain compared to the classical time sharing case.
\end{abstract}


\section{Introduction}\label{introduction}
Wireless relay networks have motivated an extremely large number of studies. When there are multiple sources, the relay might have to handle multiple access to the physical medium. If two or more sources in a wireless network transmit data at the same time, it generates interference. In first network generations, access methods strive to prevent simultaneous transmissions in order to avoid interference. Recently, the opposite approach that encourages users to interfere has been adopted. In the case of a relay network, this refers to Interference Channel with a Relay (ICR). Most strategies for ICR propose to exploit the interfered signals to increase the network capacity.

Considering an interfered signal arriving at a receiver, we focus on two mechanisms for the demodulation. Firstly, if the receiver knows a part of the interfered signal, it can perform \emph{self-interference cancellation} to subtract its own signal. Paired carrier multiple access is a practical implementation of such solution \cite{Paper:PCMA}. In two-way satellite communication systems, this technique allows two users to use the same frequency, time slot and/or code division multiple access code to transmit. Further studies of this approach have been investigated under the term Physical layer Network Coding (PNC) \cite{zhang:PNC}. Information theory demonstrates that PNC can potentially double the capacity of two-way relay network \cite{Katti:PNC}. Previous works assume a perfect synchronization in time, carrier-frequency and phase \cite{Paper:PCMA,Katti:PNC}. Asynchronous scenarios and practical deployment aspects have been studied in \cite{Zhang:PNCasynchro} and \cite{analogNC}, respectively. Secondly, if the receiver is not aware of any part of the interfered signal or has already removed its own signal, the principle is to consider the received signal as a form of \emph{superposition modulation} \cite{patent:SuperpositionCoding,superposed_mod}. These modulations result from the superposition of signals transmitted with various power levels. For instance, the authors propose to interpret pulse-amplitude modulation as the superposition of BPSK modulations with various power levels \cite{superposed_mod}. As the receiver is only interested in one part of the signal, it demodulates this part as one data stream in a hierarchical modulation \cite{hierarchical_modulation}.

In this paper, we propose a transmission scheme to increase the throughput of an ICR where $N_u$ users ($N_u \geqslant 2$) communicate through a satellite. In order to optimize the throughput, the transmission power levels are  coordinated among users. The use of satellite as relay implies low modulation orders, but our scheme can be generalized to other cases. In our scenario, each user wants to communicate with its neighbor, i.e., user $i$ transmits data to user $i+1$ (modulo $N_u$), $i=1,...,N_u$. Our scheme combines both mechanisms previously described, PNC and superposition modulation decoding. This paper has two main contributions. First, we consider the remaining signal after the self-interference cancellation as a superposition modulation. Then, we propose an evaluation of the theoretical and practical throughputs with optimal transmission power.
 
The rest of the paper is organized as follows: Section~\ref{sec:Overview} provides an overview of the proposed scheme. Section~\ref{sec:Theo} shows how to obtain the power allocations based on a capacity analysis. The performance in terms of throughput is evaluated with simulations involving Low-Density Parity-Check (LDPC) codes in Section~\ref{sec:evaluation}. Finally, we conclude the paper by summarizing the results and presenting the future work in Section~\ref{sec:Conclusion}.

\section{System Overview}\label{sec:Overview}

\subsection{Definitions and Hypotheses}\label{sec:hypotheses}
\begin{figure}[!t]
\centering
\subfigure[Time Division Multiple Access]{
\includegraphics[keepaspectratio=true, width=0.73\columnwidth]{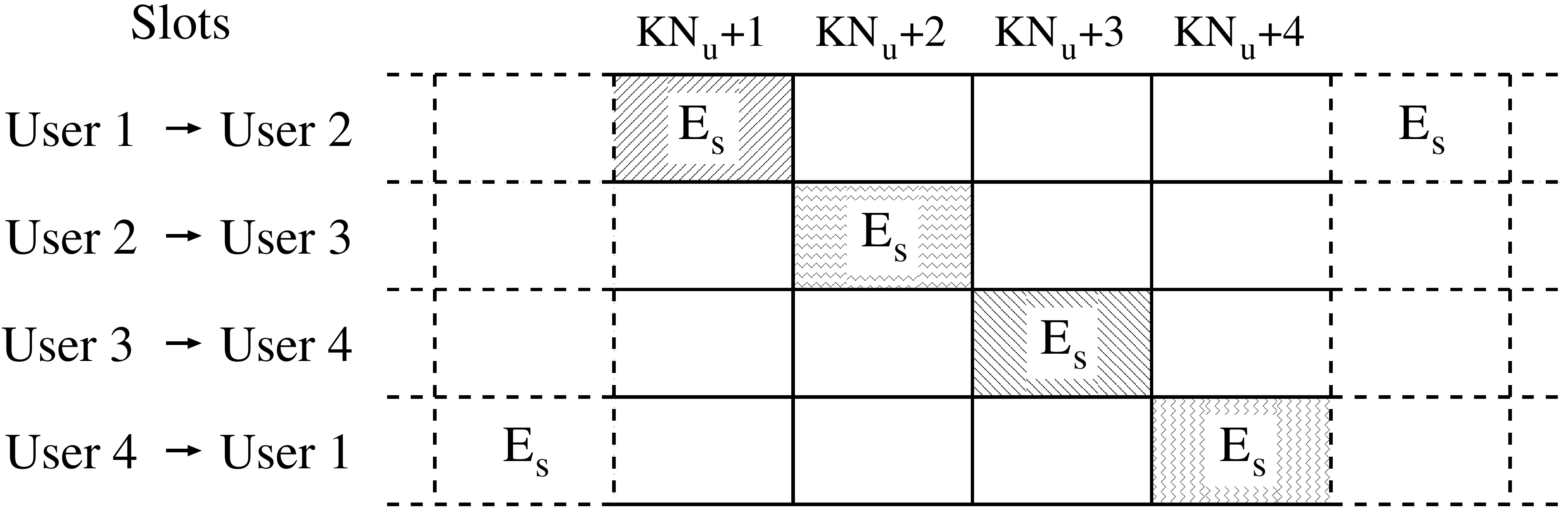}
\label{DistributionPaquetTDMA}
}
\subfigure[Proposed scheme for $N_b = 2$]{
\includegraphics[keepaspectratio=true, width=0.73\columnwidth]{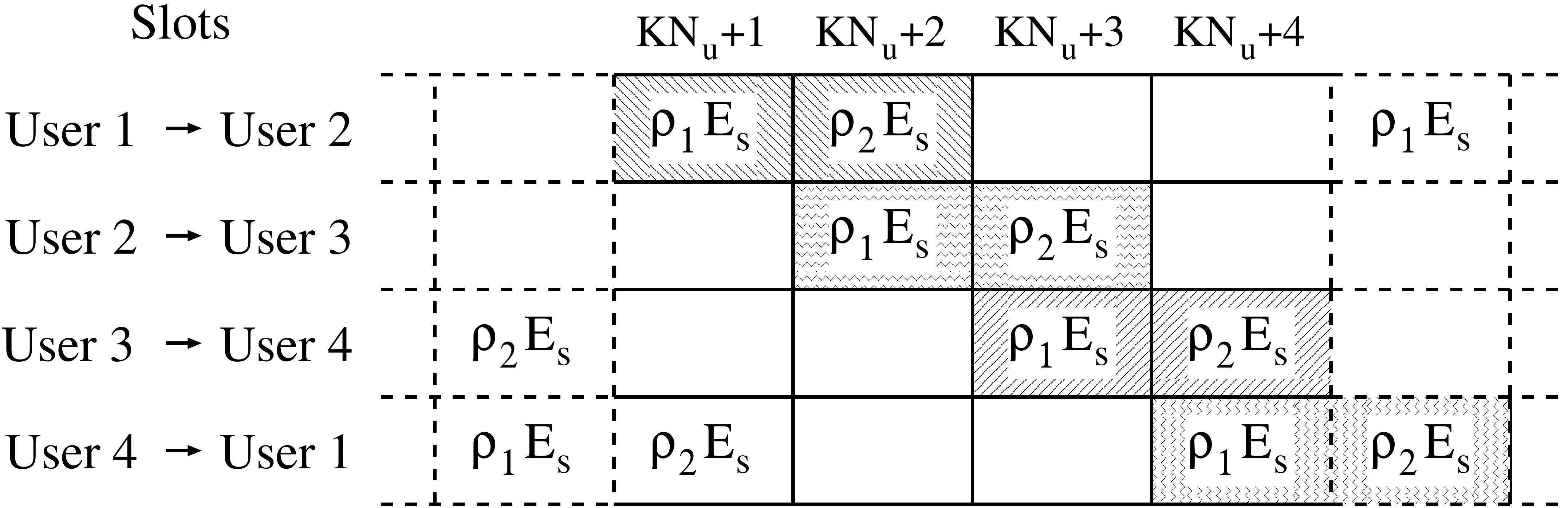}
\label{DistributionPaquet2}
}
\subfigure[Proposed scheme for $N_b = 3$]{
\includegraphics[keepaspectratio=true, width=0.73\columnwidth]{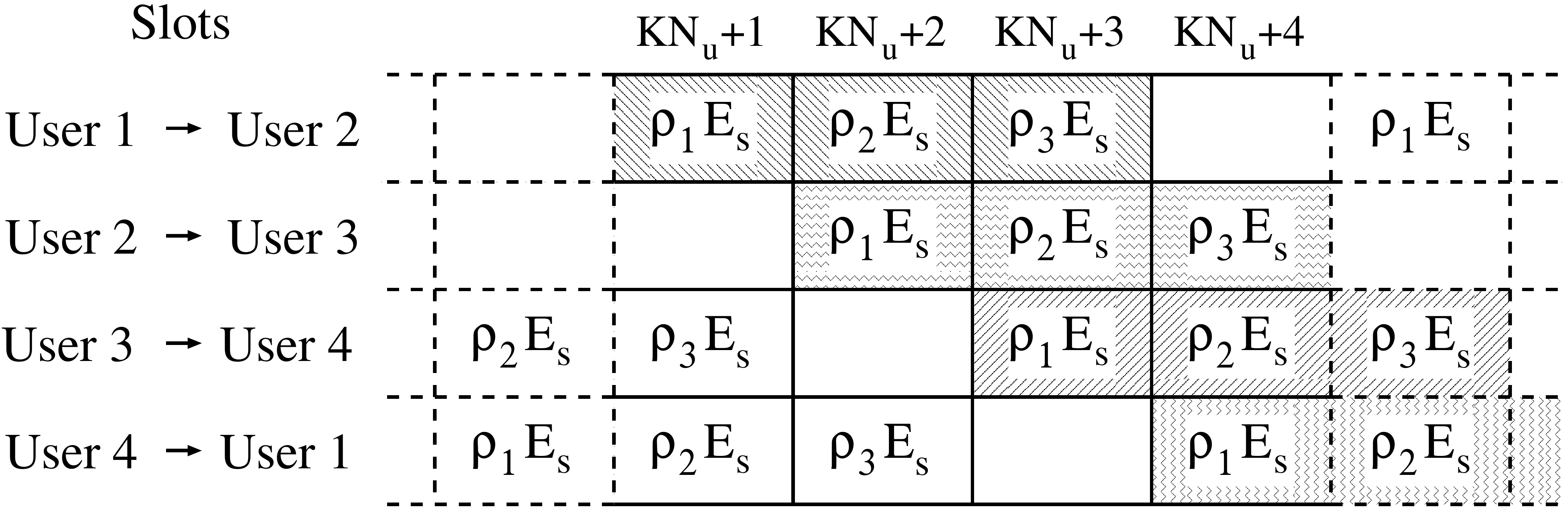}
\label{DistributionPaquet3}
}
\caption{Burst scheduling for 4 users ($N_u = 4$) on the uplink}
\label{DistributionPaquetComparaison}
\end{figure}

We consider a wireless communication system with a relay shared among $N_u$ users. The relay amplifies all received signals with a fixed gain $G$. The channel is considered linear and the transmission is subject to Additive White Gaussian Noise (AWGN). As mentioned, the relay is a satellite and each user communicates with its neighbor as shown in Figure~\ref{DistributionPaquetComparaison}. We assume that each user has the same maximum energy per symbol $E_s$ and the same link budget. Since the system aims at providing the same throughput to all users, the transmission parameters (modulation and code rate) are identical. Moreover, there is no direct link between the users. The communication medium is divided into time and/or frequency slots of same size. In each slot, we allow simultaneous transmissions. We assume that the channel estimation is perfect.

\subsection{Description of the Mechanism}\label{sec:principle}  

\subsubsection{Transmitter}\label{sec:Transmitter}
Each user transmits data packets of $k$ bits. First, an error-correcting code of rate $R$ associated with QPSK modulation is applied to these packets to create codewords of $n = k/R$ bits. Then, each codeword is split into $N_b$ physical layer packets called bursts ($N_b \leqslant N_u$). The burst size is the same for all users. Each user sends its bursts on $N_b$ consecutive slots with energy per symbol $\rho_i E_s$ in the $i$-th slot with $0 \leqslant \rho_i \leqslant 1$ ($1 \leqslant i \leqslant N_b$). More formally, user $i$ ($1 \leqslant i \leqslant N_u$) transmits its $K$-th codeword ($K \geqslant 1$) to user $i+1$ on slots number $(K-1)N_u+i$ to $(K-1)N_u+i+N_b-1$. With this scheduling, we verify that:
\begin{enumerate}
\item each user transmits a codeword on $N_b$ consecutive slots;
\item after transmitting a codeword, each user waits $N_u-N_b$ slots before sending a new one;
\item exactly $N_b$ users interfere on each slot.
\end{enumerate}   
The classical PNC scheme considered in \cite{zhang:PNC} can be seen as a particular configuration of our solution where $(N_u, N_b)=(2,2)$. The time sharing strategy, also known as Time Division Multiple Access (TDMA) (see Figure~\ref{DistributionPaquetTDMA}) corresponds to the case $N_b=1$. The cases with $(N_u, N_b) = (4,2)$ and $(N_u, N_b) = (4,3)$ are illustrated in Figure~\ref{DistributionPaquet2} and Figure~\ref{DistributionPaquet3}, respectively. We can notice that a factor $\sum_{i=1}^{N_b}\rho_i$ exists between the energies transmitted by our solution and the TDMA strategy. If $\sum_{i=1}^{N_b}\rho_i > 1$, some devices, e.g., low-power mobile devices, can suffer from this increase of global energy consumption. However, other kinds of terminal, such as very-small-aperture terminals, are limited by their maximum transmission power rather than their energy. The scheme and the assumptions considered in this paper can be then applied to this later class of terminals.

\subsubsection{Relay}\label{sec:Relay}
The relay receives a signal which is a noisy sum of signals from $N_b$ users after passing through the uplink channel. It amplifies the input signal with a fixed gain $G$ and forwards this corrupted sum of messages back to all users on a second set of time slots or on another frequency.

\subsubsection{Receiver}\label{sec:Receiver}
User $i+1$ is interested in the data transmitted by user $i$, so it only considers the signal on slots $(K-1)N_u+i$ to $(K-1)N_u+i+N_b-1$. The signal on these slots is a superposition of signals coming from multiple users after going through the channel (uplink ad downlink). In our system, the receiver has the knowledge of its own message and how this message was distorted by the channel. After correcting the channel distortion, the receiver can then subtract its message from the received signal using PNC algorithm and then infers a corrupted version of the signals of other users. This step is called self-interference cancellation. Previous study demonstrated that PNC is very robust to synchronization errors \cite{Zhang:PNCasynchro}. Thus, in the following sections, we assume that the PNC operation is perfectly done and all self-interference is totally cancelled. After the self-interference cancellation, the signals on slots $(K-1)N_u+i$ to $(K-1)N_u+i+N_b-1$ are superpositions of QPSK modulations. During the demodulation, the receiver selects the data dedicated to itself. Finally, demodulated bits from all slots are assembled and sent to the decoder. The demodulation and decoding are identical for all users.

\section{Capacity Analysis}\label{sec:Theo}

In this section, we show how to obtain the power allocations based on a capacity analysis between a sender/receiver pair. For the capacity analysis, we assume a perfect synchronization while the practical case would lead far afield \cite{asynch_capacity}. However, this assumption is not considered for the simulations in Section~\ref{sec:evaluation}. 

In our study, each user transmits a QPSK modulated signal to an amplify-and-forward relay. The signals are here subject to noise, channel attenuation and also time, phase and frequency gaps at the receiver input. We evaluate the signal-to-noise ratio (SNR) after passing through the channel between a transmitter and a receiver on any slot. In our scheme with parameters $(N_u, N_b)$, we consider the signals transmitted on the slot number $q$ ($q \ge 1$). Exactly $N_b$ users transmit on slot $q$. We denote $e_{p,q}$ the transmitted signals with an average energy per symbol of $\rho_p E_s$ ($1 \leqslant p \leqslant N_b$). The received signal on slot $q$ at the relay can be written as
\begin{equation} 
r_{relay,q}(t) = \beta_u \sum_{p=1}^{N_b} e_{p,q}(t)  + n_u(t),
\label{relayReceived}
\end{equation}
where $\beta_u$ is the path loss coefficient of the uplink channel, $n_u(t)$ is the uplink AWGN with variance $\sigma_u = 2 N_{0_u}$. The relay amplifies the input signals with a fixed gain G and forwards the sum to all users. The signal received by any user on slot $q$ is then given by
\begin{equation} 
r_{q}(t) =  \beta_d \times G \times r_{relay,q}(t) + n_d(t),
\label{DestinationReceived}
\end{equation}
where $\beta_d$ is the path loss coefficient of the downlink channel, $n_d(t)$ is the downlink AWGN with variance $\sigma_d = N_{0_d}$. The received signal SNR on slot $q$ is computed as
\begin{equation} 
SNR_q = \sum_{p=1}^{N_b}\rho_p \times \frac{E_s G^{2} \beta_u^2 \beta_d^2}{N_{0_u} \beta_d^2 G^2 + N_{0_d}}.
\label{SNRreceived}
\end{equation}

Our capacity analysis is based on superposition modulation \cite{superposition_modulation}. We define a layer as the data transmitted by a user, i.e, 2 bits per channel use. As mentioned in Section~\ref{sec:Transmitter}, each user transmits data in $N_b$ consecutive slots with power allocations $(\rho_1, ...,\rho_{N_b})$. We denote $\chi_i$ ($ 1 \leqslant i \leqslant N_b$) the corresponding constellations, i.e., QPSK constellations with energy per symbol $\rho_i E_s$. Let us consider the two constellations $\chi=\sum_i \chi_i$ and $\chi_{\smallsetminus i}=\sum_{j \neq i} \chi_j$. In our study, $N_b$ users with  energy $\rho_i E_s$ ($1 \leqslant i \leqslant N_b$) transmit on each slot, so there are exactly $N_b$ layers and each symbol of the superposition modulation carries $2 N_b$ bits. The layer $i$ corresponds to the data transmitted with energy $\rho_i E_s$. For any superposition modulation with $L$ layers, the mapping used in our work assigns the bits in positions $2l-1$ and $2l$ in the binary label of the constellation symbols to the layer with the $l$-th ($1 \leqslant l \leqslant L$) highest power.

Firstly, we compute the capacity on each slot between a transmitter/receiver pair. This capacity is similar to the capacity of one layer in a superposition modulation. For any superposition modulation $\psi$ with $L$ layers, we denote the capacity of the $l$-th layer ($l \leqslant L$) by $C^l_\psi$. An expression of $C^l_\psi$ for the AWGN case is given in \cite{hierarchical_modulation}. 

Secondly, we look for the power allocations which maximize the sum of capacities on each slot. Each user considers the signal on $N_b$ slots. After the self-interference cancellation, the receiver gets on the first slot a superposition modulation with $N_b$ layers and tries to decode the layer with energy $\rho_1 E_s$, which corresponds to the layer 1. On the $N_b-1$ remaining slots, it tries to decode one layer of a superposition modulation with $N_b-1$ layers. More formally, after cancelling its own signal with energy $\rho_i E_s$ ($1 \leqslant i \leqslant N_b-1$), the receiver observes (on the corresponding slot) the constellation $\chi_{\smallsetminus i}$. The receiver tries to decode the layer with energy $\rho_{i+1} E_s$, which corresponds to the layer $i+1$. For a given SNR between the transmitter and the receiver, the achievable rate is
\begin{equation}
R_{a}(\rho_1, ...,\rho_{N_b}) = \frac{1}{N_u} \left( C^{1}_\chi + \sum_{i=1}^{N_b-1} C^{i+1}_{\chi_{\smallsetminus i}} \right) ,
\label{capacity}
\end{equation}
where $C^{i+1}_{\chi_{\smallsetminus i}}$ corresponds to the capacity of the $(i+1)$-th layer in the superposition modulation $\chi_{\smallsetminus i}$. To achieve the theoretical rate in (\ref{capacity}), the principle is to apply a time sharing strategy with capacity-achieving codes on each slot. Note that for a practical implementation in Section~\ref{sec:evaluation}, it is preferable to use one long code with a code rate given by $1/2N_b \left( C^{1}_\chi + \sum_{i=1}^{N_b-1} C^{i+1}_{\chi_{\smallsetminus i}} \right)$. This rate corresponds to the average of the achievable rates on each slot.

The terms $C^l_\chi$ and $R_a$ depend on the SNR value and the power allocations $(\rho_1, ...,\rho_{N_b})$. For a given SNR, the power allocations are chosen in order to maximize the rate in (\ref{capacity}) and are defined as
\begin{equation}
 (\rho_1, ...,\rho_{N_b}) = \operatorname*{arg\,max}_{(x_1, ...,x_{N_b}) \in [0,1]^{N_b}} R_a(x_1,..,x_{N_b}) .
 \label{power_alloc}
\end{equation}

\begin{figure}[!t]
\centering
\includegraphics[width=0.89\columnwidth]{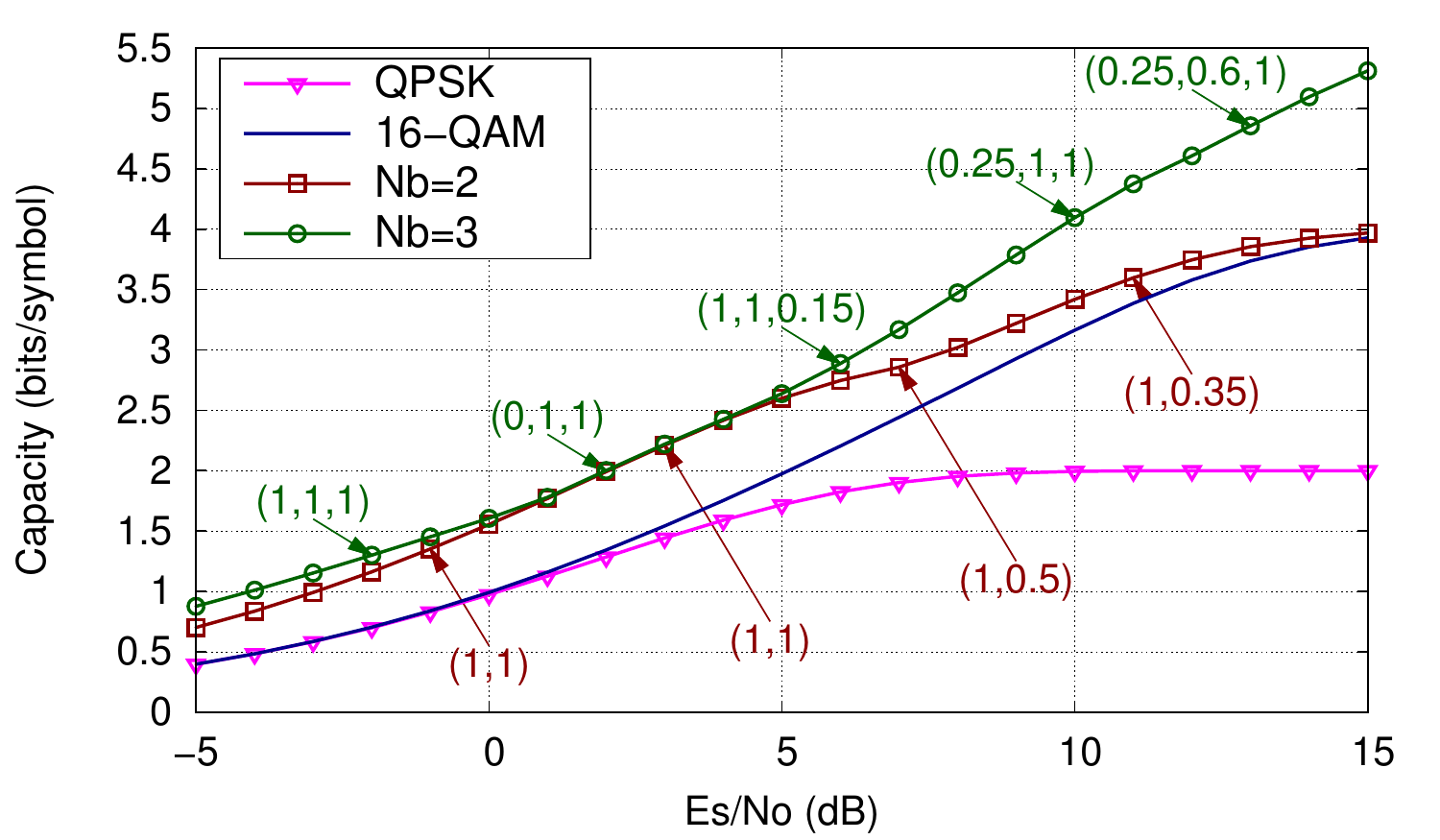}
\caption{Capacities comparison}
\label{fig:Capa}
\end{figure}

Finally, Figure~\ref{fig:Capa} shows the capacity in (\ref{capacity}) obtained with optimal power allocations for $N_b=2$, $N_b=3$ and the capacities of the QPSK and 16-QAM modulations. In the range of SNR from 0 to 5 dB, the systems with $N_b = 2$ and $N_b = 3$ obtain the same capacity. Thus, in the rest of this paper, we analyze the system with $N_b$ up to 3. This prevents to use large modulation orders as needed in satellite communications, e.g., quadrature amplitude modulation with order greater than 16 are not used in \cite{DVBS2-Standard}.  On the cooperative strategy curves, we also give the power allocations $(\rho_1,..., \rho_{N_b})$ obtained from (\ref{power_alloc}) for several SNR values. These power allocations are used for the simulations in Section~\ref{sec:evaluation}.

\section{Performance evaluation}\label{sec:evaluation}

\begin{figure*}[!ht]
\centering
\subfigure[$N_b = 2$ and $N_b=3$ vs. capacity]{
\includegraphics[width=0.44\textwidth]{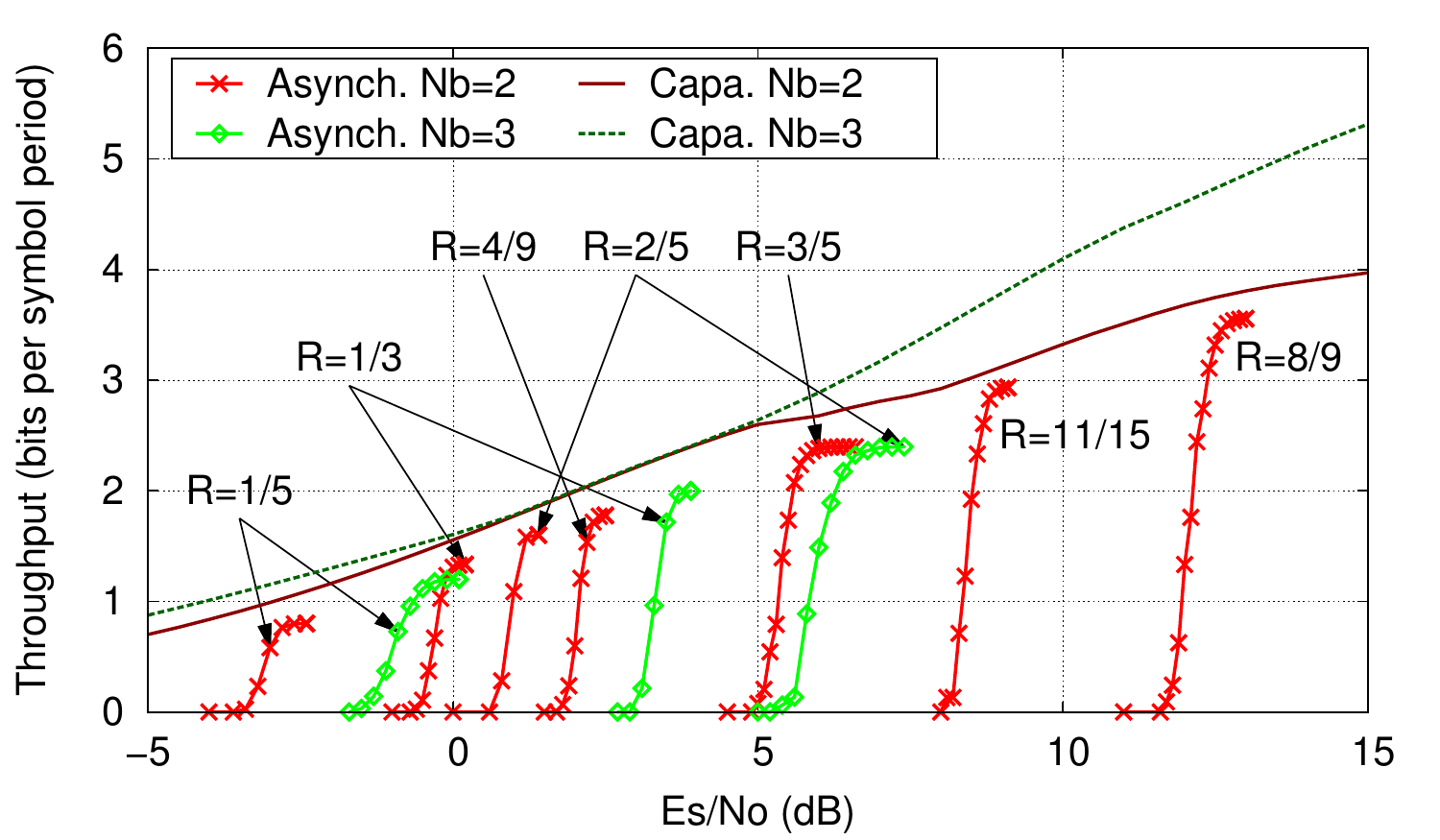}
\label{fig:SimuAsynNb2Nb3}
}
\subfigure[$N_b=2$ vs. TDMA with 16-QAM modulation]{
\includegraphics[width=0.44\textwidth]{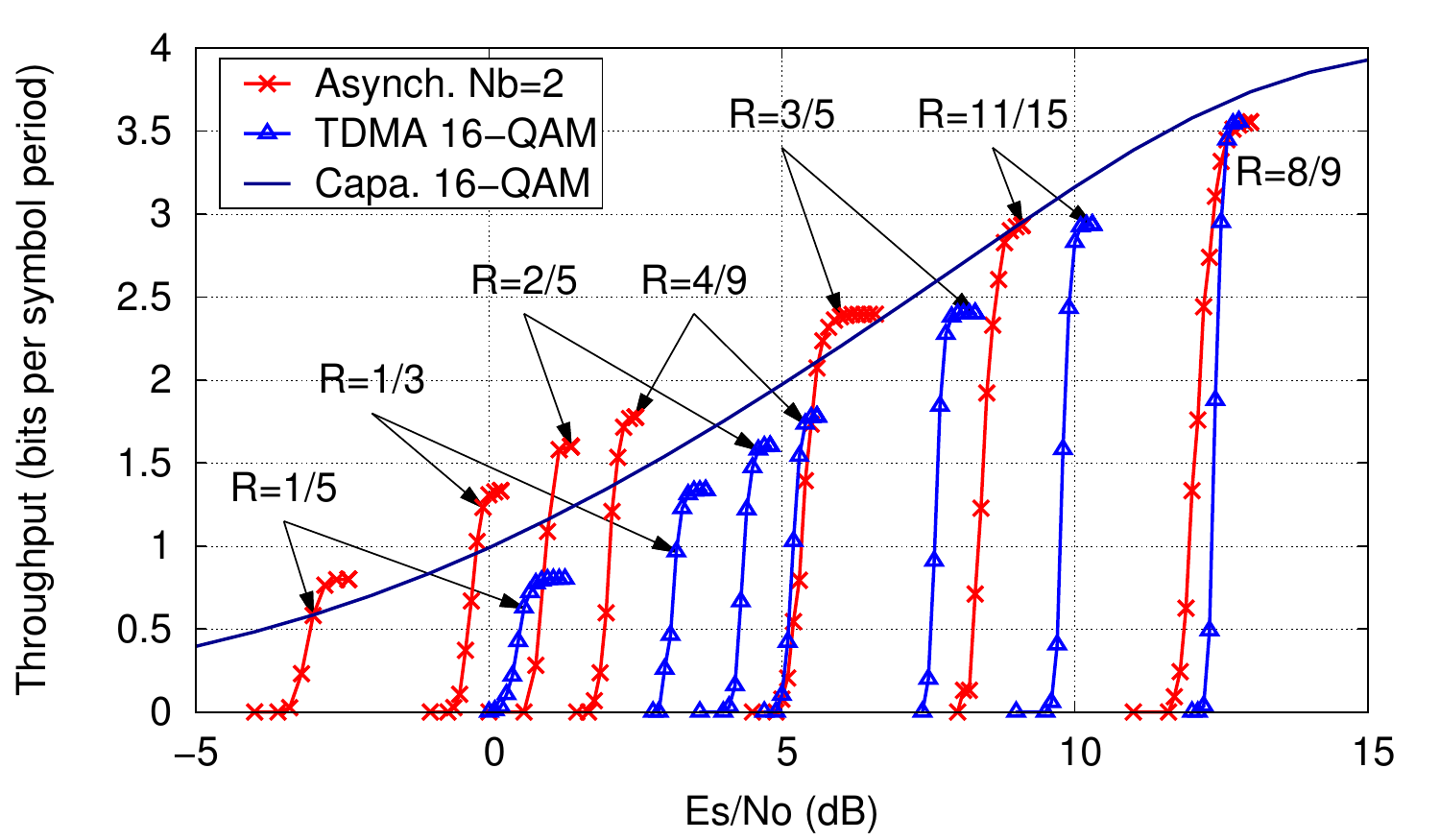}
\label{fig:SimuAsynNb2TDMA}
}
\caption{Simulations results in terms of throughput (with asynchronous assumptions for our scheme)}
\label{simu_results}
\end{figure*}
In this section, the throughput, denoted $T$, with practical error-correcting codes is studied. To realize the self-interference cancellation and the demodulation of superposed signals, we assume that the channel estimation is perfect. We also assume that all users have the same average received power. The probability of non decoding a packet, denoted PLR (Packet Loss Ratio), depends on the SNR value and the power allocations. The throughput is defined as the average number of bits successfully transmitted by the system per symbol period ($T_{sym}$). Since the codewords that contain errors after the decoding are erased, the system throughput is given by\begin{equation} 
T = \log_2(M) \times N_b \times R \times \left( 1 - PLR \right),
\label{Throughput}
\end{equation}
where $M$ is the modulation order ($M=4$ for QPSK) and $R$ is the code rate. All the data are encoded with the LDPC codes of length 16200 bits considered in the DVB-S2\footnote{Digital Video Broadcasting - Satellite - Second Generation} standard \cite{DVBS2-Standard} associated with QPSK modulation. Note that we implement a pseudo-random bit-interleaver in each codeword in order to avoid long damaged sequences at the decoder input.

In practice, it is unlikely that signals of multiple sources arrive at the destination at the exact same time with the same carrier frequency. For this reason, we study the scenarios with a lack of synchronization of few symbols and a frequency offset $\Delta f$ between interfering signals. Based on the DVB-RCS\footnote{Digital Video Broadcasting - Return Channel via Satellite} standard \cite{DVB-RCS-guidelines}, the lack of synchronization in time between two users is randomly chosen in the interval $\left[0, 4T_{sym}\right]$ and $\Delta f$ is equal to about $2\%$.

Figure~\ref{fig:SimuAsynNb2Nb3} shows the throughput according to $E_s/N_0$ when $N_b=2$ and $N_b=3$ and for several code rates. In the $N_b=2$ case, simulations show that $\rho_1 \geqslant \rho_2$ gives the best performance. Note that the throughput achieved with LDPC codes is close to the capacity. For the scenario with $N_b=3$, despite the good capacity for high SNR presented in Figure~\ref{fig:Capa}, simulations show that signals transmitted by our scheme cannot be decoded by LDPC codes with rates greater than $2/5$. This is due to the asynchronous conditions which penalize the throughput more than in the $N_b=2$ case. Thus, a throughput above 2.4 bits per symbol period cannot be achieved with the parameter $N_b = 3$. Subsequently, the parameter $N_b$ is set to 2  to keep the good performance in terms of throughput regarding to the TDMA scheme.

Figure~\ref{fig:SimuAsynNb2TDMA} shows the simulation results for our scheme with $N_b=2$ and for the TDMA scenario. The first remark is that our scheme combined with LDPC codes obtains a throughput significantly larger than the TDMA solution. Moreover, the code with rate $1/5$ combined to the parameters $(\rho_1, \rho_2) = (1, 1)$ transmits as many bits per symbol period as the TDMA case with 16-QAM modulation, but $4 \text{ dB}$ earlier. This difference vanishes when the code rate increases but it remains significant, e.g., $1$ dB for $R = 11/15$. Finally, we do not compare our solution with the TDMA scheme combined with QPSK modulation. Indeed, we see in Figure~\ref{fig:SimuAsynNb2TDMA} that our solution outperforms the 16-QAM capacity (for low SNR values) which is greater than the QPSK capacity.

\section{Conclusion and future work}\label{sec:Conclusion}

We propose a scheme based on PNC and superposition modulation decoding to increase the throughput of an ICR. Based on a capacity analysis, we show how to obtain the transmission power levels. Simulations, where imperfect synchronization in time and frequency between signals is taken into account, demonstrate a performance improvement compared to the classical TDMA scheme. Finally, our study points out that the system with $N_b=2$ gives the best performance in a satellite communication context.

In a future work, the use of other relay categories is scheduled. We also expect to investigate the impact of imperfect channel estimation on the system performance.

\section*{Acknowledgment}
This work was supported by the CNES and Thales Alenia Space.

\bibliographystyle{IEEEtran}
\bibliography{biblio}
\end{document}